\documentclass[12pt]{article}
\def \bsigma{\mbox{\boldmath $\sigma$}}
\def \btau{\mbox{\boldmath $\tau$}}

\begin{document}
\begin{center}
{\large \bf Chiral QCD, General QCD Parameterization \\ and Constituent
Quark Models}
\vskip 20 pt G.Dillon and G.Morpurgo
\end{center}
Universit\`a di Genova and Istituto Nazionale di Fisica Nucleare, Sezione
di Genova.
\footnote{e-mail: dillon@ge.infn.it \hskip0.5cm ;\hskip0.5cm
morpurgo@ge.infn.it}

\vskip 20pt \noindent {\bf Abstract.} Several recent papers -using
effective QCD
chiral Lagrangians- reproduced results obtained with the general QCD
parameterization
(GP). These include the baryon \textbf{8}+\textbf{10} mass formula, the
octet magnetic
moments and the coincidental nature of the ``perfect" $(\mu_{p}/\mu_{n})=
-(3/2)$
ratio. Although we anticipated that the GP covers the case of chiral
treatments, the
above results explicitly exemplify this fact. Also we show by the GP that
-in any
model or theory (chiral or non chiral) reproducing the results of exact
QCD- the
Franklin (Coleman Glashow) sum rule for the octet magnetic moments must
be violated.
\vskip 20pt

\noindent{\bf 1.Introduction: Chiral results and GP results} \vskip5pt
$1$- In a recent publication Durand et al.\cite{du1} (see also
\cite{du2}) derived, by
an effective chiral Lagrangian, the same \textbf{8}+\textbf{10} baryon
mass formula
that, using the general QCD parameterization (GP) \cite{m1}, \cite{mger},
had been
obtained in \cite{m.mass} and re-discussed in \cite{dm1}. Also they
derive in
Ref.\cite{du1} an expression for the baryon octet magnetic moments
similar to that
obtained with the GP in Refs.\cite{m1,mger,dm1,dmtri}. They kindly
acknowledged all
this in \cite{du1}.

$2$- These interesting results of Durand et al. add to one by Leinweber
et al.. In
Ref.\cite{lein} (see also \cite{cloet}) they show that, in a chiral
description, the
fact that the ratio $\mu(p)/\mu(n)$ is so near to $-(3/2)$ (the NR quark
model result)
is coincidental. We agree: In fact we reached the same conclusion by the
GP method
\cite{mger,dm1,dmtri} independently of a specific chiral description.

$3$- Another point discussed recently (e.g. \cite{linde}, \cite{dagu}) is
the Franklin
sum rule \cite{fr1} for the octet baryon magnetic moments. The revived
interest in the
rule is due to the fact \cite{linde} that according to a chiral quark
model, the
Franklin rule should be exact, while experimentally is not. The GP shows
that, in
fact, the rule is violated by two specific 1st order flavor breaking
terms, present in
the QCD but not in the $\chi$QM.

$4$- The GP explains the Pondrom fit \cite{po} of the magnetic moments.
\vskip 15pt

\noindent{\bf 2.The general parameterization of QCD.} \vskip5pt  The
method (GP)
\cite{m1,dm1} is derived exactly from the QCD Lagrangian exploiting only
few general
properties.\footnote{These are: 1) flavor breaking is due only to the
mass term in the
Lagrangian; 2) only quarks carry electric charge; 3) exact QCD eigenstate
can be put
in correspondence (for baryons) to a set of three quark-no gluon states
or (for
mesons) to a set of quark-antiquark-no gluon states, 4) the flavor
matrices in the
e.m. interaction and in the flavor breaking term in the QCD Lagrangian
commute.}
Although non covariant, the GP is relativistic. Also \cite{dm1} the
renormalization
point for the quark masses can be selected at will in the QCD Lagrangian.
The GP is
compatible, in particular, with a quasi-chiral Lagrangian (with light u,d
quarks) if
the latter does not violate the properties of the QCD Lagrangian. By
integrating over
the virtual $q\bar{q}$ and gluon variables, the method parameterizes
exactly the
results of the QCD calculations of various hadron properties, expressing
them in a few
body language. It allows to write, almost at first sight, the most
general expression
for the spin-flavor structure of quantities relevant to the lowest
baryons
($\bf{8}$+$\bf{10}$) and mesons. Unexpectedly one finds that, for the
lowest hadrons,
the GP is characterized, usually, by a rather small number of terms.

A consequence of the GP is that it allows to know if a constituent quark
model is
consistent with QCD. For any given property under study (masses, magnetic
moments,
etc.) it displays the exact result of QCD as a parameterized spin-flavor
expression.
The terms (and only the terms) present in the GP are compatible with QCD.

Of course solving QCD (if one could do that) would express all parameters
in the GP,
in terms of $\Lambda_{QCD}$ and the masses of the quarks. But much can be
understood
even if one is unable to calculate by QCD the values of the parameters.
This goes as
follows: Consider e.g. the masses of the lowest baryons
(\textbf{8}+\textbf{10}).
Neglecting the e.m corrections and the u, d mass difference these are 8.
But there are
only 8 GP parameters in this case; so they can be empirically determined.
After doing
this a hierarchy in the parameters emerges: The parameters multiplying
spin flavor
structures of increasing complexity are smaller and smaller. This is true
for any
quantity (not only the masses). Often this hierarchy allows to neglect
some terms in
the GP; in particular it explains why the non relativistic quark model
NRQM
\cite{nonre} works. \

\vskip15pt \noindent{\bf 3.The 8+10 mass formula. A comparison with the
chiral
results.}\vskip5pt The parameterization of the masses $M_{B}$'s of the
{\bf 8} and
{\bf 10} baryons is (for the notation and use of Eq.(\ref{pagema}) see
\cite{m.mass,dm1}; only the combination $a+b$ enters in the masses):
\begin{eqnarray}
\label{pagema} ``parameterized \ mass"=A\ +\ B\sum_i P_i^s \ +\
C\sum_{i>k}
(\bsigma_{i} \cdot \bsigma_{k}) \ + \nonumber \\
+\ D\sum_{i>k}(\bsigma_{i} \cdot \bsigma_{k})(P_i^s +P_k^s)\ +\
E\!\!\!\sum_{{\scriptstyle \begin{array}{c}i\neq k\neq
j\\(i>k)\end{array}}}\!\!
(\bsigma_{i} \cdot \bsigma_{k})P_j^s\ +\
a\sum_{i>k}P_i^s P_k^s\ + \\
+\ b\sum_{i>k}(\bsigma_{i} \cdot \bsigma_{k})P_i^s P_k^s\ +\
c\!\!\!\sum_{{\scriptstyle\begin{array}{c}i\neq k\neq
j\\(i>k)\end{array}}}\!\!
(\bsigma_{i} \cdot \bsigma_{k})(P_i^s +P_k^s)P_j^s \ +\ dP_1^sP_2^sP_3^s
\nonumber
\end{eqnarray}
In Eq.(\ref{pagema}) the flavor breaking term $\Delta m\bar \psi
P^{s}\psi$ in the QCD
Lagrangian is taken into account to all orders in $\Delta m= m_{s}-m$, no
matter how
large is $\Delta m$; thus Eq.(\ref{pagema}) includes all orders in flavor
breaking.
The values (in MeV) of the parameters in Eq.(\ref{pagema}), obtained
fitting the
baryon masses, are \cite{m.mass,dm1}:
\begin{equation}
\begin{array}{lclclcl}
\label{vapama}
A=1076 &,& B=192 &,& C=45.6 &,& D=-13.8\pm 0.3 \\
(a+b)=-16\pm 1.4 &,& E=5.1\pm 0.3 &,& c=-1.1\pm 0.7 &,& d=4\pm 3
\end{array}
\end{equation}
The hierarchy is evident. The values (\ref{vapama}) decrease rather
strongly with
increasing complexity of the accompanying spin-flavor structure so that
one can
neglect $c$ and $d$ in Eq.(\ref{pagema}) and obtain the following mass
formula
\cite{m.mass}, a generalization of the Gell-Mann Okubo formula including
octet and
decuplet:
\begin{equation}
\frac{1}{2} (p+\Xi^{0})+T=\frac{1}{4} (3\Lambda +2\Sigma^{+}-\Sigma^{0})
\label{mform}
\end{equation}
The symbols stay for the masses and $T$ is the following
combination of decuplet masses:
\begin{equation}
\label{Ti}
T=\Xi^{\ast -} - \frac{1}{2} (\Omega + \Sigma^{\ast -})
\end{equation}
Because of the accuracy reached, we wrote (\ref{mform}) so as to be free
of
electromagnetic effects before comparing it to the data. (The
combinations in
(\ref{mform}) are independent of electromagnetic and of $m_{d}-m_{u}$
effects, to zero
order in flavor breaking ($m_{s}-m$).) The data satisfy (\ref{mform}) as
follows
(using the pole values, in MeV, of the masses)
\begin{equation}
\label{strik}
l.h.s.=1133.86\pm 1.25 \qquad r.h.s.=1133.93\pm 0.04
\end{equation}
an agreement confirming the smallness of the $c, d$ terms neglected in
(\ref{pagema})
(With the conventional values of the masses the agreement is similar).

Of course \cite{dm1} a full QCD calculation, if feasible, would express
each ($A,\ B
\ldots c,\ d$) in (\ref{pagema}) in terms of the quantities in the QCD
Lagrangian, the
running quark masses and the dimensional (mass) parameter $\Lambda \equiv
\Lambda_{QCD}$; for instance: $A\equiv \Lambda \hat A(m/\Lambda
,m_{s}/\Lambda)$ where
$\hat A$ is some function. Similarly for $B,\ C,\ D,\ E,\ a,\ b,\ c,\ d$
.

\textit{The results of Durand et al.} \cite{du1} (see Sect.1) are
obtained using an
effective chiral QCD Lagrangian and heavy baryon chiral perturbation
theory. They so
reobtain our mass formula (Eq.(\ref{mform})) (our $T$ is their
$\hat{\alpha}_{MM\prime}$ in their Eq.(3.38); they did not, however,
extract the e.m.
corrections from their sum rule) \footnote{It is only after doing this
that the
agreement becomes striking, as in Eq.(\ref{strik}) above.}. Of course
this result of
\cite{du1} was expected. As stated in Sect.2, the GP is compatible with
any
relativistic chiral description, satisfying the listed general properties
of QCD (for
the pion field in the chiral Lagrangian, compare the end of Sect.4). But
it is
interesting to see this in practice, especially in view of the heavy
calculations in
the chiral treatment of \cite{du1,du2}.\footnote{In Ref.\cite{du1}(E)
``the parameter
$T$" should be read as ``the quantity $T$" (in fact $T$ is defined by
Eq.(\ref{Ti})).
Also the statement from ``so is not to be used" to ``Our approaches
differ in that
respect" is not too clear to us, because our $T$ is identical to their
$\hat{\alpha}_{MM\prime}$ (except that we included the e.m corrections).
If the above
statement means that, with the chiral effective Lagrangian, the
\textit{individual}
baryon masses can be calculated in terms of the couplings introduced in
the
Lagrangian, this may be true in principle; but, as in many chiral
treatments,
uncertainties often arise in practice.}

Clearly the work of Durand et al. also confirms indirectly the existence
of a
hierarchy, as shown by the fact that they reobtain our mass formula
(Eq.(\ref{mform}))
which is due to the smallness of $c, d$ in (\ref{vapama}).

As to the hierarchy in the GP, its field theoretical basis is discussed
in
\cite{mger}. The terms in the GP can be related to classes of Feynman
diagrams in QCD
(fig.1 of \cite{mger}). The decrease of the coefficients that multiply
terms with more
quark indices is due: a) to the increase of the number of gluons
exchanged; b) to the
fact that each flavor breaking $P_{i}^{s}$ also carries a reduction
factor. Above we
saw this for the masses. Indeed the values in Eq.(\ref{vapama}) show that
in
Eq.(\ref{pagema}) an additional pair of indices (corresponding to at
least an
additional gluon exchange between quark lines) implies a reduction factor
in the range
from 0.22 to 0.37 \cite{mger,dm1}. (One gets 0.37 using the pole values
of the
decuplet masses as we did in Eq.(\ref{vapama}) and 0.22 using the
conventional
values). The range of values 0.20 to 0.37 covers all the hadron
properties examined so
far. We will adopt usually 0.3 for the reduction factor due to ``one
gluon exchange
more". The flavor reduction factor is in the range 0.3 to 0.33. Our
reduction factor
$\approx 0.3$ is just an \textit{empirical} number derivable in principle
from QCD.
Some papers relate this $\approx 1/3$ to the $1/N_{c}$ expansion. We do
not see
similarities between the basis of the GP (an exact QCD parameterization)
and the
$1/N_{c}$ expansion (see \cite{dmarchi}).

\vskip 15pt \noindent{\bf 4.The magnetic moments of the baryon
octet}\vskip5pt To
first order in flavor breaking the parameterized magnetic moments
$M_z(B)$ of the
octet baryons $B$ derived from QCD are {\em necessarily} (notation in
Ref.
\cite{dm1}):
\begin{equation}
M_z(B)= \sum_{\nu =1}^7 \tilde g_\nu ({\bf G}_\nu)_z
\label{moct}
\end{equation}
where:
\begin{equation}
\begin{array}{c}
 \label{Ginu}  {\bf G}_1=\sum_i Q_i \bsigma_i \ ;\ {\bf G}_2=\sum_i Q_i
P_i^s
\bsigma_i \ ;\ {\bf G}_3=\sum_{i\ne k} Q_i \bsigma_k \ ;\ {\bf
G}_4=\sum_{i\ne k} Q_i
P_i^s \bsigma_k \\ \\ {\bf G}_5=\sum_{i\ne k} Q_k P_i^s \bsigma_i \ ;\
{\bf
G}_6=\sum_{i\ne k} Q_i P_k^s \bsigma_i \ ;\ {\bf G}_7 =\sum_{i\ne j\ne
k} Q_i P_j^s
\bsigma_k
\end{array}
\end{equation}
It is understood that the expectation value of the r.h.s of \ref{moct} on
the octet
spin-flavor states $W_{B}$  (compare \cite{dm1}) must be taken. Eight
${\bf G}_\nu$'s
appear in Eq.(23) of Ref.\cite{dm1}; but due to the following
Eq.(\ref{rel})-holding
for the expectation values of the ${\bf G}_\nu$'s in the $W_{b}$'s (see
\cite{dmtri})-
${\bf G}_0=Tr[QP^s]\sum_i \bsigma_i$ is expressed in terms of the ${\bf
G}_\nu$'s with
$\nu= 1...7$; thus the sum in Eq.(\ref{moct}) contains 7 terms; their
coefficients
$\tilde g_{\nu}$ differ inappreciably from the $g_{\nu}$ multiplying the
8 ${\bf
G}_\nu$. This is a very general consequence of QCD (compare the
evaluation of the
quark loop effect in \cite{dmzs0}).
\begin{equation}
\label{rel} {\bf G}_0=-\frac{1}{3} {\bf G}_1+\frac{2}{3} {\bf G}_2-
\frac{5}{6} {\bf
G}_3+ \frac{5}{3} {\bf G}_4+\frac{1}{6}{\bf G}_5+\frac{1}{6}{\bf
G}_6+\frac{2}{3}{\bf
G}_7
\end{equation}
\vskip 5pt Though the ${\bf G}_\nu$'s look non relativistic,
Eq.(\ref{moct}) is an
exact consequence of full QCD (to first order in flavor breaking). We
repeat this to
avoid misinterpreting the Eq.(\ref{moct}) as a sort of generalized NRQM.
Note the
relative dominance of $\tilde g_1$ and $\tilde g_2$ in the sum in
Eq.(\ref{moct}) (see
the Eq.(\ref{g1}) below); this is explains the fairly good 2-parameter
fit of the
naive NRQM:
\begin{equation} \label{NRQM}
\nonumber M_z(B)= \tilde g_1({\bf G_1})_z + \tilde g_2({\bf G_2})_z \quad
(NRQM)
\end{equation}
\textit{We now compare the above results with the chiral treatment of
Durand et al.
\cite{du1}}. It is notable that the latter produces precisely 7 terms for
the octet
baryon magnetic moments, to first order flavor breaking, (their Eqs.(4.6)
to (4.12))
written in terms of Pauli spin matrices, similarly to our equations
(14)(their symbols
$M$ are our $P^{s}$). Their seven \textbf{m}'s are linear combinations of
our
\textbf{G}'s. In their footnote 14 a relation appears similar to our
Eq.(\ref{rel}).

In the following we will need the magnetic moments in terms of the
$\tilde g_\nu$ 's.
>From Eq.(\ref{moct}) we get (the baryon symbol indicates the magnetic
moment):
\begin{equation}
\begin{array}{c}
\label{mfg}
p=\tilde g_1\\ \\
n=-(2/3)(\tilde g_1-\tilde g_3)\\ \\
\Lambda =-(1/3)(\tilde g_1-\tilde g_3+\tilde g_2-\tilde g_5)\\ \\
\Sigma^+=\tilde g_1+(1/9)(\tilde g_2-4\tilde g_4-4\tilde g_5+8\tilde
g_6+8\tilde g_7)\\ \\
\Sigma^-=-(1/3)(\tilde g_1+2\tilde g_3)+(1/9)(\tilde g_2-4\tilde
g_4+2\tilde g_5-4\tilde g_6-
4\tilde g_7)\\ \\
\Xi^0=-(2/3)(\tilde g_1-\tilde g_3)+(1/9)(-4\tilde g_2-2\tilde
g_4+4\tilde g_5-8\tilde g_6+10\tilde
g_7)\\ \\
\Xi^-=-(1/3)(\tilde g_1+2\tilde g_3)+(1/9)(-4\tilde g_2-2\tilde
g_4-8\tilde g_5-2\tilde g_6-
2\tilde g_7)
\end{array}
\end{equation}
and:
\begin{equation}
\label{sigla}
\mu(\Sigma \Lambda)=-(1/\sqrt{3})(\tilde g_1-\tilde g_3+\tilde g_6-\tilde
g_7)
\end{equation}
>From the PDG values of the magnetic moments \cite{PDG} we obtain:
\begin{equation}
\begin{array}{lclclcl}
\label{g1}
\tilde g_1=2.793 &;&\tilde g_2=-0.934 &;& \tilde
g_3=-0.076 &;&
\tilde g_4=0.438 \\
\tilde g_5=0.097 &;& \tilde g_6=-0.147 &;& \tilde g_7=0.154 &&
\end{array}
\end{equation}

In Eq.(\ref{g1}) the hierarchy is apparent: The average value of the one
gluon
exchange reduction factor derived from the values of  $\vert\tilde
g_{6}\vert,
\vert\tilde g_{5}\vert, \vert\tilde g_{4} \vert$ is 0.25, having adopted
$0.3$ for the
flavor reduction factor (this is $0.33$ from the ratio of $\vert \tilde
g_{2}\vert$
and $\vert\tilde g_{1}\vert$). We go on here using 0.3 for both reduction
factors;
doing so, the maximum discrepancy between estimated and empirical values
is 2.5 for
each $\vert\tilde g_{\nu}\vert$ with $\nu=4,5,6$ .

An exception is $\vert \tilde g_{3} \vert \simeq 0.08$. This is much too
small: One
expects from the hierarchy $2.79\times 0.3 \simeq 0.84$, a value 10 times
larger. We
discuss this in Sects. 5,6.\\

A comment to \textit{the Pondrom's 4-parameters fit} \cite{po} of the
baryon moments
is appropriate here. Pondrom's fit is based on the conjecture of
assigning to the
quarks different magnetic moments in different baryons and assume
additivity. But
these assumptions lead to the following approximate empirical relations
(holding to
$\pm 0.1$):
\begin{equation}
\label{er}
-(2/3)p\simeq n \hskip 30pt (1/2)(\Sigma^- -\Sigma^+)\simeq n \hskip 30pt
\Xi^- + (1/2)\Xi^0\simeq 2\Lambda
\end{equation}
The Eqs.(\ref{er}), of course, reduce from 7 to 4 the number of the
$\tilde g_\nu$'s.
One thus finds that the GP plus the smallness of $\tilde g_{3}$ explain
why a
4-parameters fit is rather good, independently of any symmetry (a
question raised in
\cite{po}). Finally, the fit \cite{po} gives $\mu (\Sigma
\Lambda)=-1.61$. The GP
formula (\ref{sigla}) - gives instead $\mu (\Sigma
\Lambda)=-1.48\pm0.04$.
Experimentally it is $\vert \mu (\Sigma
\Lambda)\vert=1.61\pm0.08$; errors are still large.\\

To go on with the comparison to the work of Durand et al.\cite{du1}, a
remark on the
pion exchange terms in a class of calculations of the baryon moments is
necessary. For
instance, for the $p$ and $n$ moments $M_{z}(p,n)$ a typical such term
is:
\begin{equation}
\label{pionex}
M_{z}(p,n)= ..............+\alpha \sum_{i\ne k}
(\bsigma_i\times\bsigma_k)_{z}(\btau_i\times\btau_k)_{3}
\end{equation}
where the dots in Eq.(\ref{pionex}) refer to the contributions other than
pion
exchange and $\alpha$ is some coefficient. Because in the exact QCD
Lagrangian only
the quark and gluon fields intervene (not those of pions), the question
arises of the
meaning of such pion exchange terms. The answer (see \cite{dmzs}) is that
they simply
duplicate terms already present in the GP; they can be always
incorporated into them.
It is:
\begin{eqnarray}
\sum_{i\ne k} (\bsigma_i\times\bsigma_k)(\btau_i\times\btau_k)_3= -8{\bf
G}_1+4{\bf
G}_3
\label{idi}
\end{eqnarray}
\textit{Durand et al. in \cite{du1} showed} that all pion exchange
magnetic moments
could be rewritten in terms of their 7 quantities \textbf{m}'s (their
Eqs.(4.6) to
(4.12) in \cite{du1}) - which are simply certain linear combinations of
our
\textbf{G}'s. We found this result (from their Eq.(4.29) to their
Eq.(4.36))
interesting also because it confirms the GP on this rather subtle point;
the pion
loops in the chiral treatment of Ref.\cite{du1} are eliminated by a
mechanism that
must be equivalent to that of Eq.(\ref{idi}).

\vskip20pt \noindent{\bf5.The coincidental nature of the ``perfect" 3/2
prediction for
\mbox{\boldmath $\vert\mu(p)/\mu(n)\vert$}}\vskip5pt
The Eq.(\ref{moct}) of the GP for
the magnetic moments, applied to $p$ and $n$, gives
\begin{equation}
\label{np}
{\bf M}(n,p)= \tilde g_{1}{\bf G}_{1}+ \tilde g_{3} {\bf G}_3 =\tilde
g_{1}
\sum_i Q_i \bsigma_i
+ \tilde g_{3}\sum_{i\ne k} Q_i \bsigma_k
\end{equation}
In Sect.4 we noted that $\tilde g_{3}$ is ten times smaller than the
value $2.79\times
0.3 \simeq 0.84$ expected from the hierarchy; the other $\tilde
g_{\nu}$'s (with
$\nu$=4,5,6) differ by no more than $2.5$ times from their expected
values. Due to
this we suggested in \cite{mger,dm1,dmtri} that the early prediction of
the NRQM
\cite{nonre} $\vert\mu(p)/\mu(n)\vert=3/2$ is coincidental.
Indeed (Eq.(\ref{mfg})) it is\\
\begin{equation}
\vert \mu(p)/\mu(n)\vert = -(3/2)[\tilde g_{1}/(\tilde g_{1}-\tilde
g_{3})]
\end{equation}
so that ($\vert \mu(p)/\mu(n)\vert $) depends critically on $\tilde
g_{3}$, the
coefficient of the second (non additive) term in Eq.(\ref{np}).

Recently Leinweber et al. \cite{lein} reached the same conclusion, that
the almost
perfect 3/2 prediction is coincidental. In Ref.\cite{lein}
$\vert\mu(p)/\mu(n)\vert$
is calculated in a chiral QCD perturbation theory, dynamically broken by
pions; the
above ratio varies from 1.37 to 1.55 as the pion mass varies from 0 to
$\simeq 280$
MeV (corresponding to ``a variation of current quark mass from 0 to just
20 MeV").
\textit{Again, as with the work of Durand et al., the chiral conclusion
agrees with
that from the GP}. We try however to clarify some statements in
Ref.\cite{lein} and in
Cloet et al. \cite{cloet}: 1) The assertion in Ref.\cite{lein} that
``within the
constituent quark model the ratio $\vert\mu(p)/\mu(n)\vert$ would remain
constant at
3/2, independent of the change of the quark mass" is correct only in an
additive
model, such as the original NRQM \cite{nonre}. The GP expression
(\ref{np}) for the
$p, n$ magnetic moments in a constituent model obtained from QCD, shows
that this is
not additive (for the importance of non additivity see also
Ref.(\cite{hoga}). 2)
Cloet et al. \cite{cloet}, referring to the GP as ``something a little
more
sophisticated than the simplest constituent quark model", add a statement
on ``the
need to incorporate meson cloud effects into conventional constituent
quark models".
There is no doubt that constituent quarks must be dressed, but this was
already there
in the old additive NRQM \cite{nonre}. The GP incorporates all $q\bar{q}$
and gluon
effects \cite{m1}; in particular the magnetic moments of the {\bf8}
baryons in the GP
are those of \textit{any possible} constituent quark model compatible
with QCD,
endowed with the correct ``dressing" of the (constituent) quarks.

To conclude: The chiral Lagrangians of Durand et al. and of Leinweber et
al. produce
results in agreement with the GP. As to the question: What about the pion
field that
appears in the chiral Lagrangians of \cite{du1} and \cite{lein}, but not
in the exact
QCD Lagrangian (where the pion is not an independent field), this has
been answered in
Sect.4 for the magnetic moments and a similar argument should be true for
the masses
(as shown by the results of \cite{du1}).
\vskip15pt \noindent{\bf 6.Parameterizing the magnetic moments of the
decuplet:
$\mathbf{p,n,\Delta}$} \vskip5pt We apply now the GP to the magnetic
moments of the
$\Delta$'s in addition to $p,n$. This  clarifies further the mechanism
producing
accidentally a small value of $\tilde g_{3}$ (noted in the past section)
and thus the
coincidental nature of $\vert \mu(p)/\mu(n) \vert = 3/2$. Also we obtain
some results
on the $\Delta$'s. The general QCD spin-flavor structure of the magnetic
moments of
$p, n, \Delta'$s is \cite{m1,dmtri}:
\begin{equation}
\label{mudel} \mu (B)= \sum_{perm} [\alpha Q_{1}+\delta (Q_{2}+Q_{3})]
\bsigma_{1z} +
[\beta Q_{1}+\gamma (Q_{2}+Q_{3})]\bsigma_{1z}
(\bsigma_{2}\cdot\bsigma_{3})\
\end{equation}
The Eq.(\ref{mudel}) is the same as Eq.(62) of \cite{m1};
$\alpha,\delta,\beta,\gamma$
are four real parameters. The sum over perm(utations) in (\ref{mudel})
means that to
the term (123) one adds (321) and (231).\footnote{In \cite{m1} correct
the following
misprints: In Eq.(63) insert $(-2\gamma)$ in the second square brackets;
in Eq.(66)
write F=$\delta -\beta -4 \gamma$; in Eq.(64) (Q term) replace $-2\gamma$
with
$+4\gamma$.}

We adopt the "standard" hierarchy for the parameters
$\alpha,\delta,\beta,\gamma$ with
the reduction factor 0.3 for one more gluon exchange; but because this
factor for the
magnetic moments is between
 $0.2$ and $0.3$ -see the remarks in Sect.4 after
Eq.(\ref{g1})- we widen the error in $\delta $, the largest
parameter after the dominant one $\alpha$ \\
\begin{equation}
\label{valdel}
\hspace{2cm}\vert\delta/\alpha\vert = 0.2\leftrightarrow 0.3;
\hspace{2cm}\vert \beta/\delta\vert \approx 0.3;
\hspace{2cm}\vert \gamma/\delta \vert \approx 0.3\\
\end{equation}
\noindent  From Eqs.(\ref{mudel},\ref{np}) we obtain for $\tilde g_{1}$
and $\tilde
g_{3}$
\begin{equation}
\label{kkk}
\tilde g_1=\alpha -3\beta -2\gamma \hspace{15pt};\hspace{15pt} \tilde
g_3=\delta -\beta-4\gamma
\end{equation}
 From Eq.(\ref{np}) one has for the the magnetic moments ($p,n$), now
indicated by the particle symbols, expressed in proton magnetons:\\
\begin{equation}
\label{mu}
p = (\alpha - 3\beta -2\gamma)\hspace{1cm} ,
\hspace{1cm} n = -(2/3)(\alpha - \delta -2\beta +2\gamma)\
\end{equation}
Hence:
\begin{equation}
\label{n/p}
(n/p)= -2/3(1+ [-\delta +\beta +4\gamma]/p)
\end{equation}
Thus the deviation of $\vert n/p \vert$ from (2/3) is determined by the
term in square
brackets in Eq.(\ref{n/p}). If it were not for the \textit{second order
terms} $(\beta
+4\gamma)$ with $\beta$ and $\gamma$ of order $(0.3)^{2}$ (three
indices), the
dominant deviation would be of order $\vert \delta/p \vert = 0.25\pm
0.05$; that is
20\% to 30\% of the ``perfect" value 2/3. To summarize, the mechanism
giving to
$\vert(p/n)\vert$ a value so near to (3/2) is this: In Eq.(\ref{n/p})
$(\beta +
4\gamma)$ almost cancels $(-\delta)$ (which is $> 0$), producing,
accidentally,
$(n/p)$= -(3/2) to a few percent.
One can also show that $\beta$ and $\gamma$ must have
opposite signs and $\gamma$ is $<0$. This is derived using the $(\Delta
\to
p\gamma)_{0}$ matrix element extrapolated to vanishing transferred photon
momentum
($k=0$) that we know experimentally. We do not enter on this here
(compare
\cite{dmtri} where, however, some data must be changed). Here we just
give the
approximate values of $\alpha,\delta ,\beta , \gamma$. It is: $\alpha
\simeq 3$ and
the values of $\delta,\beta,\gamma$ (affected by the errors stated in
Eq.(\ref{valdel})- are: $\delta= -0.75, \beta= 0.25, \gamma= -0.25$.

One more point: From Eq.(\ref{mudel}) one can also express in terms of
$\alpha,\delta,\beta,\gamma$ the magnetic moments $\mu(\Delta)$ of the
$\Delta$'s. It
is:
\begin{equation}
\label{muDelta}
\mu (\Delta ) = (\alpha +2\delta +\beta +2\gamma )Q_{\Delta }= (\mu (p)
+2\delta
+4\beta +4\gamma)Q_{\Delta}
\end{equation}

In view of the above, the magnetic moment of the singly charged
$\Delta^{+}$,
$\mu(\Delta^{+})$ (the coefficient of $J_{z}$), is expected to be
appreciably smaller
than $\mu(p)$, but the error on its value is large. (Of course it is
$\mu(\Delta^{++})
\simeq 2\mu(\Delta^{+})$. We stress that, in general, it is
$\mu(\Delta^{Q})$ = kQ +
$\xi$, but $\xi$ is negligible \cite{dmzs0} because it is a Trace term
strongly
depressed by the exchange of several gluons needed by the Furry
theorem.\vskip 15pt

\noindent \textbf{7.Some remarks on a sum rule for the baryon octet
magnetic
moments}\vskip5pt \indent Long ago Franklin \cite{fr1} suggested a sum
rule for the
baryon octet
magnetic moments that is often called the Coleman-Glashow rule since
it has the same form as the Coleman-Glashow rule for the electromagnetic
mass differences. Franklin's rule
is:\\
\begin{equation}
\label{FCG}
\Sigma{_{\mu}}= 0
\end{equation}
where it is
\begin{equation}
\label{Linde}
\Sigma{_{\mu}} \equiv
\mu(p)-\mu(n)+\mu(\Sigma^{-})-\mu(\Sigma^{+})+\mu(\Xi^{0})-\mu(\Xi^{-})
\end{equation}
Because in  chiral models of baryons of the Manohar-Georgi type
($\chi$QM)
the rule of Eq.(\ref{FCG})
should be satisfied \cite{linde}, but in reality it is violated:\\
\begin{equation}
\label{049}
\Sigma{_{\mu}}= 0.49\pm 0.03
\end{equation}
some work has been done \cite{linde,dagu} to understand the reason for
this fact.
After showing that in a ($\chi$QM) the rule (\ref{FCG}) is satisfied,
Linde et al.
consider several extensions of the $\chi$QM (several phenomenological
models) that
break the rule. We refer to \cite{linde,dagu} for many references and a
description of
the models.

Here we
will only show that it follows from the GP that the rule is necessarily
broken
by two specific first order flavor breaking terms, $\tilde g_{5}$ and
$\tilde g_{6}$.
Indeed, using the Eqs.(\ref{mfg}), (\ref{g1}) it is:
\begin{equation}
\label{coef}
\Sigma{_{\mu}}\equiv 2(\tilde g_{5} -\tilde g_{6})
\end{equation}
With the $\tilde g_{\nu}$'s given in Eq.(\ref{g1})
the r.h.s. of Eq.(\ref{coef})
is in fact $(0.49\pm 0.03)$.

The interest of the above deduction stays in its conclusion: \\
Any Lagrangian or phenomenological model (chiral or non chiral) designed
to reproduce
the result of the exact QCD Lagrangian violates the Franklin
(Coleman-Glashow) sum
rule for the octet baryon magnetic moments \textit{if, and only if,} the
coefficients
of the flavor breaking terms ${\bf G}_{5}$ and ${\bf G}_{6}$  do not
vanish and their
difference does not vanish. In such case the model must be built in such
a way that
2$(g_{5}-g_{6}$) is equal to
the experimental value 0.49$\pm$0.03. All other parameters,
multiplying ${\bf G}_{\nu}$ with $\nu \not= 5,6$, even the flavor
breaking ones with
$\nu=2, 4, 7$, do not produce violations of the rule. \vskip20pt
\noindent
\textbf{8.Conclusion}\vskip5pt \noindent In Ref.(\cite{dm1}) we stated
that the
general QCD parameterization (GP) explains why a large variety of
different theories
and models, \textit{including relativistic chiral theories}, may work
successfully.
Now, thanks especially to the treatment by Durand et al. \cite{du1} of
the baryon
masses and magnetic moments, we have an explicit detailed confirmation
that the GP
covers (see also \cite{lein}) the case of the relativistic chiral field
theories,
provided that such theories (or models) are compatible with the general
properties of
QCD.

\newpage

\end{document}